\documentclass[preprint,aps]{revtex4}
\begin{document}
\preprint{BAS-01}
\title{\LARGE\bf The application of computational mechanics to the analysis of 
geomagnetic data}
\author{Richard~W.~Clarke}
\altaffiliation{Now at Queens' College, Cambridge, U.K. }
\email{rwc25@cantab.net}
\author{Mervyn~P.~Freeman}
\author{Nicholas~W.~Watkins}
\affiliation{British Antarctic Survey, Madingley Road, Cambridge, U.K.}
\date{\today}
\begin{abstract}
We discuss how the ideal formalism of Computational Mechanics can be 
adapted to apply to a non-infinite series of corrupted and correlated data, that
 is typical of most observed natural time series. Specifically, a simple filter
  that removes the corruption that creates rare unphysical causal states is 
  demonstrated, and the new concept of effective soficity is introduced. We believe
   that Computational Mechanics cannot be applied to a noisy and finite data 
   series without invoking an argument based upon effective soficity. A related 
   distinction between noise and randomness is also defined: Noise can only be
    eliminated by increasing the length of the time series, whereas the
     resolution of random structure only requires the finite memory of the
      analysis to be increased. The benefits of these new concepts are 
      demonstrated on simulated times series by (a) the effective elimination 
      of white noise corruption from a periodic signal using the expletive
       filter and (b) the appearance of an effectively sofic region in the 
       statistical complexity of a biased Poisson switch time series that is 
       insensitive to changes in the wordlength (memory) used in the analysis. 
       The new algorithm is then applied to analysis of a real geomagnetic 
       time series measured at Halley, Antarctica. Two principal components in 
       the structure are detected that are interpreted as the diurnal variation 
       due to the rotation of the earth-based station under an electrical
        current pattern that is fixed with respect to the sun-earth axis and 
        the random occurrence of a signature likely to be that of the magnetic 
        substorm. In conclusion, a hypothesis is advanced about model 
        construction in general.
        \end{abstract}

        \maketitle
        
\section{Introduction}
\label{Introduction}
Computational mechanics (CM) \cite{Badii1999} has a formalism 
\cite{Crutchfield1989} that has been proved to construct the minimal model 
capable of statistically reproducing all the resolvable causal structure of any 
infinite sequence of discrete measurements (be they scalar, vector, tensor, or 
descriptive)\cite{Shalizi2001}. The size of a model so defined, measured by a 
quantity termed statistical complexity, $C_{\mu}$, \cite{Crutchfield1989} is a 
reliable and falsifiable indication of the amount of structure the data 
contain\cite{Crutchfield1994, Shalizi2000}.

The particular strengths of this approach are that it enables the complexities 
and structures of different sets of data to be quantifiably compared and that it 
directly discovers detailed causal structure within those data. By examining 
data in this way it is possible to appreciate, in a well-defined abstract sense, 
how a system actually functions and what scales are most important to it. This 
information can then be used to optimise the efficiency of physically plausible 
models\cite{Palmer2000}.

As with all other analytical tools, CM has some limitations in the face of 
certain real-world problems that affect the information content of the signal 
under study.  These problems may include:

\begin{enumerate}
\item{Gaps in the data}
\item{Noise}
\item{Restricted sequence length}
\item{Correlations at a very wide range of scales.}
\end{enumerate}

The problem of correlations at a wide range of scales is particularly 
interesting and relevant to geophysical and other natural time series because of 
their typically power law (coloured noise) Fourier spectra \cite{pt1999}.  
Theoretically, the minimum resolvable scale will be constrained by the data 
sampling interval and the maximum resolvable scale by the length of the data 
series.  In practice, the range of resolvable scales will also be set by the 
available computational resources. Thus, the range of resolvable scales may be 
less than those necessary to evaluate correlations on all relevant scales.  
Consequently, it is important to understand how structural analysis is affected 
by unresolved structure due to correlation.

In this paper, we address these issues in detail.  In section~\ref{Method}, we 
discuss how the ideal formalism of CM can be adapted to apply to a non-infinite 
series of corrupted and correlated data.  In particular, three concepts are 
defined and discussed:  (1) A tolerance parameter \cite{Crutchfield1989} to 
account for the statistical uncertainty introduced by a non-infinite series that 
destroys the exact equivalence of different causal states sharing the same 
outcome.  (2) A new expletive filter that removes signal corruption by assuming 
that corruption creates rare causal states or words that are not in the 
dictionary of the true signal.  (3) The new concept of effective soficity in 
which a data series has a finite set of equivalent causal states that is stable 
to small changes in the effective memory of those states.

The latter concept distinguishes between unresolvable ``random" structure and 
resolvable structure whose discovery is only prevented by the effective memory 
being used and by the length of the data series.

In section~\ref{Examples}, we apply the CM algorithm with these additional 
concepts to the analysis of structure in four simulated time series:  (1) 
Uncorrelated, white noise.  (2) Periodic signal with white noise corruption.  
(3) A biased Poisson switch (i.e., a sequence of pulses whose pulse durations 
and inter-pulse intervals are determined by stationary Poisson processes).  (4) 
A sequence of bursts similar to (3) but with fixed pulse duration.  The 
structure of the time series is analysed by searching for regions of effective 
soficity in maps of statistical complexity over the parameter space of the CM 
model.

The simulated time series represent four types of signal thought to be present 
in time series measurements of the geomagnetic field.  In 
section~\ref{Application}, we use the CM algorithm to examine a real geomagnetic 
time series measured at Halley, Antarctica, in which deflections of the earth's 
magnetic field are due mainly to electrical currents in the ionosphere.  The CM 
analysis yields a structural model that comprises a diurnal component 
corresponding to the oscillation of the measuring apparatus with the rotation of 
the earth and a Poisson-switched, fixed-duration, pulse component that is likely 
associated with the magnetospheric substorm \cite{Borovsky1993}.

In section~\ref{Discussion}, we discuss some general principles that have been 
learnt in applying CM to the analysis of structure in real data, and draw 
conclusions in section~\ref{Conclusion}.

\section{Method}
\label{Method}
Here we give an introduction to the practical use of CM in the analysis of real 
data. We concentrate only on describing in detail the formalism for the parsing 
structure that we have used in the analyses. For a fuller description of the 
potential intricacies of the method see reference\cite{Shalizi2001}. Defining 
some new terminology, we highlight the difficulties associated with analysing 
experimental data in this way, and explain solutions to these problems.

To start with, one has a set of measurements - either a spatial or temporal 
series where the separation between each point is known. The total time or 
length for which data exist is their span, $S$. After coarse-graining at a fixed 
scale $s$, the series has $N = \frac{S}{s}$ equally spaced measurements. Next, 
we digitise the signal amplitude. For reasons that will be apparent later, the 
number of possible digits should be low unless the series length is extremely 
large. The digitised sequence is then a concatenation of $N$ letters 
${\bf{l}}^{N} = \{l_{0},l_{1},\ldots,l_{i},l_{N-1}\}$, where there are $L$ types 
of such letters, ranging from $0,1,2,\ldots$ up to $L-1$. In order to maximize 
the prior probable information content of the processed sequence, digitisation 
should normally be performed such that there are equal numbers of each letter 
present. For example, in the case of binarisation (where $L = 2$), this would 
mean that the threshold for letter $1$ would be the median value of the data. It 
should be noted, though, that the best way to digitise the sequence is that 
which {\em actually} maximises the information content of the result; but that 
this cannot usually be guessed. Another approach that has been 
suggested\cite{Palmer2000} is to use the formalism of Maximum Entropy.

The next step is to parse the sequence. One begins by composing words from each 
group of $n$ consecutive letters; the $i$th word, ${\bf W}_{i}$, is defined by:
\begin{equation}
{\bf W}_{i} = {\bf l}_{i}^{n} = \{l_{i},l_{i+1},\ldots,l_{i+n-1}\}
\end{equation}
Thus there are $L^{n}$ possible words, each represented by a unique scalar, 
$W_{i}$:
\begin{equation}
W_{i} =  \sum_{j=0}^{n-1}L^{(n-1)-j}\times l_{i+j}
\end{equation}
The total number of words generated from the sample is $N-(n-1)$. We now 
introduce some terminology; any word ${\bf W}_{i}$ may be called a {\em 
proword}, ${\bf W}_{P}$ when followed by any word ${\bf W}_{i+1}$. This latter 
is called the {\em epiword}, ${\bf W}_{E}$. For this sentence we digress 
slightly to note that it may sometimes be beneficial to perform the initial 
digitisation on each separate block of data $2n$ letters long rather than the 
entire dataset.

We now proceed to capture causal structure in the word sequence by compiling a 
tally of epiwords following each proword. This means going through the sequence 
incrementing an array ${\bf T}_{(W_{P},W_{E})}$ accordingly. Representing 
summation over an index by its omission, we see that the total tally is ${T} = 
N-(2n-1)$. Thus, contracting over epiwords gives a tally of prowords only:
\begin{equation}
{\bf T}_{(W_{P})}=\sum_{W_{E}=0}^{L^{n}-1}{\bf T}_{(W_{P},W_{E})}
\end{equation}
and the fractional prevalence of each proword in the sequence is therefore 
contained in the vector
\begin{equation}
{\bf P}_{(W_{P})} = \frac{{\bf T}_{(W_{P})}}{T}
\end{equation}
Finally, the fractional {\em profile} of each proword by epiword is given by the 
array
\begin{equation}
{\bf P}_{(W_{E}\mid W_{P})}= \frac{{\bf T}_{(W_{P},W_{E})}}{{\bf  T}_{(W_{P})}}
\end{equation}
where the repeated indices in the division are not summed over. Given a 
particular proword, this tells us the likelihoods of transitions to the various 
epiwords.

The crux of the technique now lies in identifying prowords with equivalent 
epiword profiles. Such prowords are said to belong to the same ``equivalence 
class'' or ``causal state'' - i.e. they share statistically equivalent 
probabilistic futures (at the level of analysis one has been pursuing). The 
identification is made via an equivalence relation, denoted by $\sim$. For an 
infinite sequence, $\sim$ can demand exact correspondence between profiles, in 
which case it is always transitive (meaning $A\sim B, B\sim C  \Rightarrow A\sim 
C$). In a practical situation, where even the finite length of the sequence 
introduces fluctuations in the calculated profiles\cite{footnote1}, it is not 
possible to be so exact. We therefore introduce a tolerance parameter, $\tau$, 
within the bounds of which the profiles of words in the same equivalence class 
are allowed to vary: Two prowords, $A$ and $B$, are in the same equivalence 
class if, ${\bf\forall}$ $W_{E}$;
\begin{equation}
\left|{\bf P}_{(W_{E}\mid W_{P}=A)}-{\bf P}_{(W_{E}\mid W_{P}=B)}\right| \leq 
\tau
\end{equation}
where the large vertical bars signify absolute magnitude. 

Although this appears to destroy the formal transitive property of $\sim$, it is 
postulated here that after some finite sequence length is surpassed, there 
always exists a definite range of values for $\tau$ within which the transitive 
property is observed correctly. The transitive property can be enforced where 
$\sim$ is not transitive --- by grouping equivalence classes defined by $\sim$ 
that share at least one word.

Having identified the words lying within each equivalence class, a model which 
outputs a series of letters statistically equivalent to the original can be 
constructed. It is a particular strength of the technique that the model 
generated is always a minimal representation of the data's statistical structure 
for the amount of memory the analysis employs\cite{Shalizi2000}. The model is 
easiest to describe in terms of its representation as a labelled ``diagraph". 
Two very simple labelled diagraphs, with extracts from their outputs, are 
presented in Figure~\ref{TwZrCmplxtDgrphs}. A more complicated labelled 
diagraph, also representing a minimal model, is shown in 
Figure~\ref{MrCmplctdDgrph}. Each diagraph comprises a node or nodes, indicated 
by a circle with a number in it, and lines joining one node to another or to 
itself. Each numbered node of a diagraph represents a causal state corresponding 
to each of the model's equivalence classes, while each line (unidirectionally) 
joining two nodes is labelled with the string of letters (the word) that is 
output when that line is followed. In addition, each line is associated with a 
probability. The word output on going from one causal state to another is in the 
equivalence class of the future state. The probability of each word's output may 
therefore be trivially given by ${\bf P}_{(W_{E}\mid W_{P})}$. It should be held 
in mind that only a subset of all possible labelled diagraphs represent minimal 
models. Even so, an arbitrary diagraph's output can naturally be used to 
construct the appropriate minimal model.
\newline
Models like this are useful for three reasons:
\begin{enumerate}
\item Their minimality allows the structure of two sets of data to be directly 
compared.
\item Once a model has been synchronized with current data it optimizes one's 
ability to forecast the behaviour of the system in the future.
\item The information concerning scales of causal structure in the data can be 
used to optimize the performance of more physically plausible models.   
\end{enumerate}
If a recursive decomposition is 
employed\cite{Zurek1990,LamNaroditsky1992,Crutchfield1994,Hanson1993,Perry1999,
Shalizi2000,Young1991}, diagraphs labelled with words to output can be reformed 
into equivalent diagraphs labelled with single letters. This is perhaps a 
mathematically aesthetic thing to do. However, because any real analysis is 
performed with an effective memory of only $n$ symbols, only the last $n$ 
symbols are of any use for prediction. This fact is not manifest in the 
single-letter diagraph obtained through the decomposition of the transition 
matrix, $P_{(W_{E}\mid W_{P})}$. Moreover, it is necessary to synchronise a 
single-letter diagraph to an input stream of data before it is of any use for 
the purpose of prediction. In this paper we concentrate only upon the 
identification of proword equivalence classes, because it is a powerful tool for 
pattern discovery in its own right.

A measure of the structure of such models is given by the statistical 
complexity\cite{Crutchfield1989}:
\begin{equation}
C_{\mu} \equiv -\sum_{i}P(C_{i})\log_{2}P(C_{i})
\end{equation}
where logarithms are canonically taken to base $2$ and the prevalence 
$P_{(C_{i})}$ of equivalence class $i$ is given by the sum of the prevalences of 
the words in that class.
For example, the models represented by the labelled diagraphs in 
Figure~\ref{TwZrCmplxtDgrphs} both have a statistical complexity of zero because 
they only have one causal state (and therefore one equivalence class) each. This 
is sensible because they both output noise. The model represented in 
Figure~\ref{MrCmplctdDgrph}, though, has four causal states with equal 
prevalences and a correspondingly higher statistical complexity of two bits:
\begin{equation}
C_{\mu} = -\sum_{i=1}^{4}P(C_{i})\log_{2}P(C_{i}) = -4\times \frac{1}{4}\times 
\frac{\ln(\frac{1}{4})}{\ln(2)} = 2
\end{equation}

$C_{\mu}$ is extremely important, not only because it reflects the complexity of 
the system, but also because it does not converge until the data have been fully 
characterized. It is a hard fact that if the sequence length $N$ is too small, 
full characterization will not be possible. This is because fluctuations in the 
proword profiles will corrupt the identification of equivalence classes. In this 
text the resultant unresolvable structure is called {\em noise}. In contrast, 
resolvable but as yet unresolved structure is described as {\em random} 
\cite{Poirier2001}. Such random structure is likely to be encountered in 
analysing data sets with correlation lengths comparable to or exceeding the 
maximum wordlength. Making this distinction is very important, even though it is 
not possible to discern whether unresolved structure is noisy or random until 
further computation has resolved it. In other words, the data appear to be noisy 
until the series is found to be {\em effectively sofic}, at which point 
$C_{\mu}$ attains its correct value and the model is complete. We note that 
sofic sequences are those which still have a finite number of equivalence 
classes when $N$ is infinite and $n$ is semi-infinite (see Badii and 
Politi\cite{Badii1999}, page 80 for a longer explanation). {\em Effective} 
soficity is here defined to mean that a sequence has equivalence classes that 
are stable to an increase in wordlength. Thus, a sequence could be effectively 
sofic at one range of wordlengths but not at another where either more or less 
structure is in the process of being identified.

Both random structure and noise will redistribute the original tally from what 
would be expected if only resolved structure was present, raising $C_{\mu}$ from 
the value corresponding to resolved structure alone and increasing the 
complexity of its model. A simple one-parameter model for the corruption process 
is to assume that the probability that any letter is corrupted to any other 
letter is $\chi$. Then the probability any letter stays as it is is 
$\sigma=1-\chi$ and the corruption will have been governed by the redistribution 
function
\begin{equation}
{\bf T}_{(W_{P}^{C},W_{E}^{C})}^{\mbox{corrupt}} = \sum_{W_{P}=0}^{L^{n}-1} 
\sum_{W_{E}=0}^{L^{n}-1}P(\${\bf W}_{P}={\bf W}_{P}^{C},\${\bf W}_{E}={\bf 
W}_{E}^{C}){\bf T}_{(W_{P},W_{E})}^{\mbox{pure}}
\end{equation}
where $\${\bf A}={\bf B}^{C}$ reads, {\em the pure word ${\bf A}$, when 
corrupted by noise in a certain way, is identical to the corrupt word ${\bf 
B}^{C}$}. The label ``pure'' implies effective soficity. Thus, assuming the 
corruption of prowords and epiwords are independent we have
\begin{equation}
P(\$ {\bf W}_{P}={\bf W}_{P}^{C},\$ {\bf W}_{E}={\bf W}_{E}^{C}) = P(\${\bf 
W}_{P}={\bf W}_{P}^{C})P(\$ {\bf W}_{E}={\bf W}_{E}^{C})
\end{equation}
where
\begin{equation}
P(\$ {\bf W}={\bf W}^{C})=
\prod_{i=1}^{n}\left\{ \sigma\delta(w_{i}= w_{i}^{c})+\chi\delta(w_{i}\neq 
w_{i}^{c}) \right\}
\end{equation}
where $\delta$s is a Kronecker delta and the corruption of letters is assumed to 
be independent.

It happens that arbitrarily corrupted distributions can be uniquely deconvolved 
as long as one knows $\chi$, but this is not usually the case in an experimental 
situation. We have two alternative options. The first is to scan through $\chi$, 
deconvolving the proword prevalences each time. This will produce a drastic 
decrease in the statistical complexity at some point, signifying correct 
parameterization of $\chi$. A good guess for $\chi$ might be the first value 
which results in a single proword having a prevalence of zero.

Whilst the assumption of independent corruption of letters is likely to be a 
good model of noise, it is unlikely to be a good model of the uncharacterised 
correlated structure that we call random. Consequently, a second option is to 
ignore the details of any corruption and simply assume that the prevalence of 
any expletive (corrupted word) is below a certain expletive prevalence, $x$. We 
scan through $x$, eradicating any prowords whose prevalence is less than $x$, 
and recalculate $C_{\mu}$ each time. This procedure can alternatively be 
performed after the identification of preliminary equivalence classes to 
eradicate expletive equivalence classes. In any case, the approach can only work 
when the actual structure-to-noise ratio (SNR) is high enough to ensure that 
expletives are eradicated before meaningful words are. If the pure proword 
prevalence distribution is very uneven this method cannot work. In general, a 
combined method would probably be most successful - that is, where one first 
attempts the deconvolution after making some bold assumptions and then removes 
the resulting low-prevalence words completely. It is always possible to 
determine all the resolvable structure of a sequence for which the SNR is 
arbitrarily small, so $C_{\mu}$ is independent of SNR. Of course though, if the 
SNR is zero, so is $C_{\mu}$, because the model suddenly collapses to a single 
equivalence class.

Note that deconvolution can always be achieved by inversion of an assumed 
convolution matrix, but that this is not always easy. In particular, if one knew 
the actual matrix then the ``noise'' would not be noise at all, but resolved 
structure. The only deconvolution that is strictly necessary is that which 
removes the noise (unresolvable structure) from the signal. It should therefore 
assume that the redistribution is Gaussian. In practice though, some random 
(resolvable) structure may be so computationally difficult to identify that a 
messy deconvolution is required to remove it, allowing the analysis of more 
easily resolvable structure to proceed. It is admissible to remove expletives 
from the prevalence distribution because they destroy the effective soficity of 
the data.

It is instructive at this point to go through the uncertainties present in the 
profile and prevalence distributions. When the sequence length is large compared 
to $L^{n}$ the probability that any individual word has been corrupted is 
approximately $\Delta =n\chi$. Following the definition of the prevalence 
distribution, we find that the uncertainty in the prevalence of a proword 
$\Delta P_{(W_{P})}$ is governed by an inequality:
\begin{equation}
\frac{\Delta}{\sqrt{T}} < \Delta P_{(W_{P})} < \Delta
\end{equation}
where the lower limit corresponds to uncorrelated errors and the upper limit to 
systematic errors. We indeed expect the uncertainty to be somewhere in this 
range because the errors are due to unresolved structure. The uncertainty in the 
prevalence of a single epiword within a particular proword's profile is expected 
to be greater:
\begin{equation}
\frac{\Delta}{\sqrt{{\bf T}_{(W_{P})}}} < \Delta P_{(W_{E}\mid W_{P})} < \Delta
\label{uncertainty}
\end{equation}
These inequalities go some way to justifying the use of the blanket tolerance 
$\tau$ to identify the equivalence classes, because we know nothing about the 
nature of the errors. In some cases it is conceivable that $\tau$ would have to 
be scaled by $1/{{{\bf T}_{(W_{P})}}^{m}}$, where $0<m<\frac{1}{2}$  in order to 
correctly identify equivalence relations between profiles. In such cases $m$ is 
an extra parameter. 

\section{Examples}
\label{Examples}
We now turn to the analysis of test sets of data by the algorithm described in 
detail above.  The test data represent signal types thought to be present in 
time series measurements of the geomagnetic field that we shall study in the 
next section.

\subsection{White noise.}
A white noise (temporally uncorrelated) signal was generated by a sequence of 
5000 independent samples from a uniform distribution and converted to binary by 
setting those values above the median to unity and those below the median to 
zero.  Figure~\ref{RndmSttstclCmplxt} shows the variation of statistical 
complexity, $C_{\mu}$, versus word length, $n$, and tolerance, $\tau$, for this 
signal. The absence of a plateau in this graph indicates that, for the range of 
memories (wordlengths) tested, the analysis does not discern any structure at 
all in the signal. The linear variation of $C_{\mu}$ with $n$ for $\tau \approx 
0$ represents models with as much arbitrariness as is possible at each level of 
memory used in the analysis. These models collapse to a single equivalence class 
as the tolerance parameter is increased. An increasing amount of tolerance is 
required for this collapse for increasing wordlength, as expected from 
equation~\ref{uncertainty}. Thus, no complex models at all were constructed for 
this noisy sequence at any time during this analysis. This was expected; we 
would have been disappointed with the random number generator that was used to 
construct the sequence (the IDL ``randomu" function, see also\cite{Press1996}) 
if we had easily found correlations.

\subsection{Periodic signal with white noise corruption.}
Figure~\ref{10PrcntNsBnrPrd4SttstcclCmplxt} shows the result of the analysis on 
a binary period four sequence (i.e., 00110011...) of length 5000, where 10\% of 
the bits have been randomly flipped. This graph has a stable, but rather jagged, 
plateau at $C_{\mu} \approx 2.8$ which begins at wordlength 4 for tolerances in 
the range $0.1 \leq \tau \leq 0.24$. This plateau corresponds to a group of 
models that capture the essential structure in the signal. In the absence of 
noise the statistical complexity of a binary period four signal should be 
$C_{\mu} = 2$. The apparently anomalously high level of the plateau is caused by 
both the noise and the finite sequence length corrupting the identification of 
the equivalence classes. It is not entirely flat because the corruption is 
different at each value of wordlength and tolerance. In fact, there is a gentle 
downward trend which would converge to $C_{\mu}=2$ in the limit of the extra, 
spurious, states decreasing in prevalence at longer and longer wordlengths, if 
the sequence was long enough. Note that the gradient of the increase of 
statistical complexity with wordlength changes at a memory equal to half the 
period of the structure in this signal. This is the point at which the structure 
is first discovered: It is important to note that the convergence of $C_{\mu}$ 
is not immediate, suggesting an analogue of the Nyquist sampling theorem for CM. 
Note also steep drops in $C_{\mu}$ where previously distinguishable equivalence 
classes have suddenly collapsed together as the tolerance parameter $\tau$ 
exceeds some critical value, supporting our earlier postulate that there is a 
definite range of $\tau$ within which a true transitive property is observed.

Figure~\ref{10PrcntNsBnrPrd4SttstcclCmplxtX075} shows results of a similar 
analysis on the same sequence, excepting that this time, words of prevalence 
less than $x$ (the expletive prevalence parameter) were eradicated from the 
probability distributions. $x$ was chosen to be 0.075 for this graph. As we can 
see, this approach was entirely successful in the respect that $C_{\mu}$ 
converges to a plateau for a broad range of the tolerance parameter $\tau$. A 
minimal model that was capable of outputting sequences with statistical 
structure identical to that characterized from the input was effectively 
constructed at every point on this plateau. The value of $C_{\mu}$ for periodic 
sequences was always found to directly reflect the amount of memory required by 
the system to produce such data: A sequence with a sole period, $Q$, has a 
statistical complexity of $\log_{2}(Q) \mbox{ bits}$ (in this case the period 
four signal has a statistical complexity of $2.0 \mbox{ bits}$). Moreover, we 
can appreciate that the analysis only yields a convergent value after the 
wordlength has exceeded at least half the period of the sequence. More 
generally; convergence begins when an analysis first has greater memory than a 
system. If the system has certain structure with greater memory than it may be 
feasible to analyse, for example, a red noise signal that consists of many 
Fourier modes with a power law distribution of amplitudes and random phases, 
$C_{\mu}$ will not ever truly converge. However, there may be stages where the 
analysis has enough memory to identify {\em some} structure, and this is 
indicated by approximately flat regions, or, at the very least, dips in the 
gradient of $C_{\mu}$ with increasing wordlength.
\subsection{Biased Poisson switch.}
We next turned to more detailed analyses of two other illustratively important 
diagraph's outputs. The first we considered was the biased Poisson 
switch\cite{Bendat1958}, represented as a labelled diagraph in 
Figure~\ref{PssnSwtchDgrph}. The circled states, $0$ and $1$ may each generate 
either a one or a zero with the probabilities shown.

In the figure, $\bar{\alpha}=1-\alpha$ and $\bar{\beta}=1-\beta$. Note that when 
$\alpha=\beta$ the output sequence is no longer biased. It turns out that the 
values of $C_{\mu}$ we can derive for different values of $\alpha$ and $\beta$ 
provide some nice insights into the nature of information and the optimization 
of measurement processes. The measure has two distinct regimes: where 
$\alpha+\beta=1$, and where they do not. Since the diagraph only has two states, 
it is clear that as far prediction of the next epiword is concerned, only the 
last bit of any proword can ever matter. Therefore, all words usually separate 
into two equivalence classes (corresponding to odd and even words). If, however, 
$\alpha+\beta=1$ then $\alpha=1-\beta=\bar{\beta}$ and $\beta=\bar{\alpha}$. 
This always results in the two equivalence classes collapsing into one, giving a 
statistical complexity of zero, corresponding to pure noise. This is appropriate 
because in this degenerate situation the possible outcomes of node 0 in 
Figure~\ref{PssnSwtchDgrph} are identical to those of node 1 and the diagraph 
collapses to a single state too (see inset), and can only produce noise anyway. 
If the diagraph does not collapse in this way there will always be two 
equivalence classes. Their prevalences are found to be 
$1/(\frac{\alpha}{\beta}+1)$ and $1/(\frac{\beta}{\alpha}+1)$. Thus, when 
$\alpha +\beta \neq 1$, we have:
\begin{eqnarray}
C_{\mu} &=& -\sum_{i}P_{i}\log_{2}\left( P_{i}\right) \nonumber \\
&=& \frac{1}{\left[ \frac{\alpha}{\beta}+1\right] }\log_{2}\left( 
\frac{\alpha}{\beta}+1\right) + 
\frac{1}{\left[ \frac{\beta}{\alpha}+1\right] }\log_{2}\left( 
\frac{\beta}{\alpha}+1\right)
\end{eqnarray}
and if $\alpha +\beta =1$, $C_{\mu}$ is always zero. A graph of this function is 
shown in Figure~\ref{PssnSwtchSttstcclCmplxtVrssBs1}. Note that it always 
evaluates to unity when $\alpha =\beta$, except when $\alpha$ = $\beta 
=\frac{1}{2}$. If $\alpha$ does not equal $\beta$ (and $\alpha +\beta \neq 1$) 
then it is less than unity. In fact, as the switch becomes more and more biased 
the statistical complexity goes down and down, reaching zero when only one digit 
is ever output. This is to be expected because a biased data set (e.g. more ones 
than zeros) is a symptom of an inefficient measurement apparatus: If one symbol 
is more prevalent than any other then the system is under-characterised by the 
alphabet in use. In the parlance of Shannon's theory of communication this 
statistical complexity is equivalent to the maximum rate of information.

Given that the collapse discussed above takes a slice out of the graph in 
Figure~\ref{PssnSwtchSttstcclCmplxtVrssBs1}, we would expect sequences generated 
by certain Poisson switches to be more difficult to characterize. For example, 
sequences produced by a switch with $\alpha =\beta =0.49$ have a statistical 
complexity of unity, but it is difficult to distinguish them from noise (where 
$C_{\mu}=0$) because they are so close to the collapse at $\alpha =\beta 
=\frac{1}{2}$. Such a sequence, 5000 symbols long, was analysed up to a 
wordlength of $n=7$ at 100 equal intervals between $\tau=0.00$ and $\tau=0.01$, 
without assuming any noise was present (i.e. $\chi =0$ and $x=0$). The results 
are presented in Figure~\ref{AB49PssnSwtchSttstcclCmplxt}. The plateau 
corresponding to the optimal model is that which has a statistical complexity of 
unity. We can see that it is difficult to construct this model because the 
plateau is radically constricted at higher wordlengths. On one side of it $\tau$ 
is too small to identify the equivalence classes, so every proword occupies its 
own equivalence class and $C_{\mu}=n$, its maximum value at any wordlength $n$. 
On the other side, $\tau$ is too large, so the two equivalence classes collapse 
together, producing degenerate models that would output noise. The $C_{\mu}=1$ 
plateau has a distinct end at $n=5$ because the sequence is not long enough to 
support analysis at a wordlength of $n=6$. At the latter wordlength statistical 
fluctuations in every proword profile mean that the correct classification of 
equivalence classes is no longer possible at any range of $\tau$. We are not too 
concerned about this here because we have already identified the optimal model 
which was stable from $n=1$ to $n=5$. In fact, a ``more optimal'' model would be 
able to predict the flipping of the switch itself to some extent. The 
construction of such a model would probably need a lot of computation and would 
probably require $N$ to be very large. These things depend on how random or 
noisy the switch is.

If a set of data is very complicated, no stable model might be identified before 
the wordlength becomes too large to be statistically supportable by the sequence 
length. The only solution is to gather more data. The alternative is to settle 
with models that are either inadequate or arbitrarily complicated. Although the 
latter models reproduce structure well (and are therefore most useful to 
engineers), studying them can reveal little about underlying processes. They are 
scientifically unaesthetic. In contrast, one can tell a lot about the 
intricacies of a system from the minimal adequate model associated with it at a 
certain level of analysis. This is the concern of scientists.

\subsection{Fixed pulse duration Poisson switch.}
The next class of labelled diagraphs we consider produce binary sequences that 
are simple models of a process with bursts. These sequences have the structure 
of sustained switches - that is, when the switch is down it has a constant 
probability of switching up, and when up, it stays up for a fixed count, $U$. 
When the sequence is unbiased the up-switching probability is $\frac{1}{U+1}$. 
See Figure~\ref{4SmblSstndSwtchDgrph} for an example of this kind of labelled 
diagraph. The exact statistical complexities of such unbiased sustained switches 
are given by
\begin{eqnarray}
C_{\mu} &=& -\left [ \left ( \frac{U+1}{2U}\right )\log_{2}\left (\frac{U+1}{2U} 
\right) + 
(U-1)\left(\frac{1}{2U}\right)\log_{2}\left(\frac{1}{2U}\right)\right] \nonumber 
\\
&=& \frac{1}{\ln(2)}\left[\ln(2U)-\left(\frac{U+1}{2U}\right)\ln(U+1)\right]
\end{eqnarray}

We now investigate the practical analysis of a sequence one million binary 
symbols long that was produced by a sustained switch with $U=4$. The statistical 
complexities of the models constructed by the analysis are shown in 
Figure~\ref{4SmblSstndSwtchSttstcclCmplxt}. It can be seen that the first 
convergent values are at wordlengths one greater than $U$. That is to say, good 
models can be constructed when the analysis first has a greater memory than the 
system. The plateau identifiable with a model of the form shown in 
Figure~\ref{4SmblSstndSwtchDgrph} begins at a wordlength of $4$ and extends 
laterally from $\tau\approx0.02$ to $\tau\approx0.06$. The remarkable thing 
about this plateau is that, although it is very flat, it is not {\em entirely} 
flat. It begins at $C_{\mu}^{5}\approx1.77069$ which is significantly higher 
than the theoretical statistical complexity of:
$  C_{\mu}=\frac{1}{\ln(2)}
\left[\ln(8)\left(\frac{5}{8}\right)\ln(5)\right]\approx1.54879  $
and subsequently oscillates around this value while it converges to 
it  (e.g. $ C_{\mu}^{10}\approx1.52160 $). This behaviour is caused by the phase 
ambiguity due to the absence of information concerning the synchronisation of a 
burst when a word is composed entirely of ``up'' symbols. For example, at 
wordlength six, the profile of word $63$, (i.e., $111111$ in binary), is a 
superposition of the profiles of sequences like $10[111111]$, $01[111111]$ and 
$11[111111]$, from each of which it cannot be distinguished at that level of 
analysis. Therefore, in this case, the profile of word $63$ does not match that 
of any other word, and is allocated its own equivalence class. Although the 
prevalence of this word, and thence its class, is very low, it is sufficient to 
distort the statistical complexity.

In an analysis with recourse to infinite memory, the prevalence of an infinite 
sequence of ``up'' symbols is zero. Thus, the $U$ causal states of such a 
sequence would be correctly identified, and the statistical complexity of the 
model constructed would match exactly with the theoretical value. Of course, in 
practice no analysis can have infinite memory. If one wishes to retain optimal 
predictability of future data then it is necessary to accept whatever model is 
actually constructed by an analysis with finite memory.

\section{Analysing geomagnetic data}
\label{Application}
The test data examples analysed in the previous section represent signal types 
thought to be present in time series measurements of the geomagnetic field.  If 
this is true, we may expect to see similar structure emerging from a CM analysis 
of a real geomagnetic time series.

The CM analysis detailed in section~\ref{Method} was performed on 3-hour 
averaged measurements of the variation of the east-west component of the 
geomagnetic field at Halley, Antarctica, from three separate years: 24 February 
- 16 December, 1995, 26 January - 28 December, 1998, and 2 January - 30 December 
2000.  A graph of the data from 26 January to 28 December 1998 is shown in 
Figure~\ref{MgntcDflctn1998}.  It can be seen that the magnetic deflections have 
both a linear trend and a high frequency signal with an annual amplitude 
modulation that maximises in the austral summer.  The linear trend is caused by 
the movement of the ice shelf upon which Halley is situated and was removed by 
subtracting the result of a linear regression for each of the three years.  The 
detrended time series was then binarised with respect to the median, giving 
three sequences of 2352, 2688, and 2896 symbols, respectively. These series were 
then analysed up to a wordlength of 10 and with tolerances varying in 80 equal 
steps from 0.05 to 0.25. Words with prevalences less than $x = 0.004$ were 
eradicated.  The graph of statistical complexity, $C_{\mu}$, is shown in 
Figure~\ref{180MntMgntcDflctnSttstcclCmplxt}.  Two plateaus are evident, one at 
$C_{\mu} \approx 0.9$, covering a wide range of tolerances and between word 
lengths of 1 and 3, and the other plateau at $C_{\mu} \approx 5.0$, at the top 
left-hand corner of the graph, between tolerances of about 0.05 and 0.07 and at 
word lengths of 8 or more.

The convergence of statistical complexity at a word length of 8 corresponds to a 
time scale of 8 x 3 = 24 hours.  Such a diurnal variation is well known and is 
primarily caused by the rotation of the observing station with the earth under 
the so-called SQ ionospheric current system that is driven by pressure gradients 
caused by solar heating and is thus fixed in the Sun-Earth frame 
\cite{Akasofu1972}.  The variation can be seen in the raw data, as illustrated 
by plotting a typical month of Halley geomagnetic data in 
Figure~\ref{MgntcDflctn199802}.  The associated ground magnetic variation has 
neither a pure sinusoidal shape nor a fixed period of exactly 24 hours, that is 
likely to account for the higher observed statistical complexity of $C_{\mu} 
\approx 5.0$ compared to the expected $C_{\mu} = 3$ for a pure binary period 8 
signal.

The other plateau in Figure~\ref{180MntMgntcDflctnSttstcclCmplxt} at word 
lengths of 1 to 3 indicates the presence of significant structure at 3 to 9~h 
time scales.  This plateau has a statistical complexity of approximately 0.9 and 
an overall structure similar to that of 
Figure~\ref{AB49PssnSwtchSttstcclCmplxt}, suggesting the possibility of some 
random pulse-like process.  Such a possibility is intriguing because pulse-like 
geomagnetic perturbations on hour time scales (known as magnetic bays) are 
particularly prominent during the nighttime at high (auroral zone) latitudes and 
are associated with magnetospheric substorms \cite{Smith1999} whose occurrence 
has been argued to be a stationary Poisson process with mean recurrence time of 
5 h \cite{Borovsky1993}.  Figure~\ref{MgntcDflctn19980616} shows a single day of 
Halley geomagnetic data that illustrates the presence of such pulse-like 
disturbances on hour time scales sitting on top of the diurnal variation.

To investigate this further, an analysis was made of a 40-minute averaged time 
series of the east-west component of the geomagnetic field at Halley from 00:00 
UT, 25 January, 1998 to 00:00 UT, 26 December, 1998.  After removing the linear 
trend in the data due to the movement of the ice shelf, the time series was 
binarised with respect to the median, giving a sequence of 12011 symbols.  The 
series was analysed up to a word length of 11 for tolerances in 60 equal steps 
between 0.00 and 0.15.  Words with prevalences less than $x = 0.015$ were 
eradicated.  The graph obtained for $C_{\mu}$ is shown in 
Figure~\ref{40MntMgntcDflctnSttstcclCmplxt}. The plateaus in this graph are 
stable to variation of $x$. The higher plateaus have models that are more useful 
for prediction of future data, if they are stable to an increase in the amount 
of data available to the analysis. The lower plateaus have models that show the 
most dominant structures -- and are easier to understand and interpret 
physically.

It can be seen from the graph that, at a tolerance between $\tau = 0.12$ and 
$\tau = 0.15$, more structure is identified between wordlengths six and eight 
than it was possible to resolve with a memory of only five symbols. The model 
which corresponds to this plateau is represented, for a wordlength of seven, in 
Figure~\ref{40MntMgntcDflctnDgrph}. The details of the model are in the 
Appendix. Comparing with Figure~\ref{PssnSwtchDgrph}, the transitions between 
states 0 and 1 of this diagraph are an approximately Poisson-switched process 
with a timescale of about five hours. This value is given by the range of 
wordlengths capable of resolving this structure from the sequence within this 
range of $\tau$ ($n = 6, 7 \mbox{and} 8$); at $n=7$ the characteristic timescale 
is $7\times40\mbox{ minutes} \approx 4\mbox{ hours and }40\mbox{ minutes}$.

It was thought that the other states and transitions in 
Figure~\ref{40MntMgntcDflctnDgrph} would be caused by the diurnal variation of 
the data alone. This was investigated by analysing, in exactly the same way, a 
pure binary sequence with a period of 36 symbols -- corresponding to a period of 
one day if each symbol were to represent a 40-minute average. The principal 
transitions of the model constructed for this sequence are shown in the diagraph 
drawn in Figure~\ref{Prd36Dgrph}. The structural similarities and differences 
between this diagraph and the one in Figure~\ref{40MntMgntcDflctnDgrph} are 
obvious, and support the idea that the transitions between states 0 and 1 of 
Figure~\ref{40MntMgntcDflctnDgrph} are due to substorm activity, rather than 
merely being an artifact of a partially characterised 24-hour period.

\section{Discussion}
\label{Discussion}
In the previous sections, we have demonstrated how CM can measure the 
statistical complexity of linear data sequences and construct the minimal model 
necessary to describe the data. The reader may have noticed that there are seven 
degrees of freedom in making such a model:
\begin{enumerate}
\item digitisation method (binary, trinary, etc)
\item coarse-graining scale, $s$
\item sequence length, $L$
\item wordlength, $n$
\item tolerance, $\tau$
\item corruption frequency, $\chi$
\item expletive frequency, $x$
\end{enumerate}

These degrees of freedom express the level of information in the data and the 
depth of knowledge with which the model is probing the system from which the 
data are measured. For example, increasing the sequence length, $L$, reducing 
the coarse-graining scale, $s$, or increasing the digitisation from binary to 
trinary, all provide increased information and thereby increased knowledge of 
the system that the data represent. Conversely, increasing the tolerance or the 
expletive frequency reduces information by admitting different states to be 
equivalent or to be omitted, respectively, thereby reducing knowledge of the 
system. Consequently, we might anticipate that the best model of the system is 
the model corresponding to the region of the multi-dimensional parameter space 
in which information is maximised. Whilst such a model is the most accurate 
description of the data sequence with the greatest information content, it is 
not necessarily the optimal model of the system. This is because any data 
sequence is not a complete representation  of the system it is measured from. In 
particular, it is limited in two important respects: First, there is structure 
in a data sequence, that we have termed noise, that cannot be resolved under any 
amount of computation. This will create differences in the profiles of words 
that are statistically insignificant and should be ignored by allowing some 
non-zero value of tolerance, corruption frequency or expletive frequency. 
Second, there is structure in a data sequence, that we have termed random, that 
has not been resolved at a certain level memory or wordlength but that is 
resolvable at a greater wordlength. Recognising these sources of structure, we 
advance a hypothesis about model construction:
\begin{quote}
Given enough data relevant to a system, there exists at least one connected 
region of finite size in the space of parameters used to construct models from 
those data within which an optimal model of the system is constructed.
\end{quote}

In other words, meaningful models of the data can only be found within certain, 
usually finite, zones of the parameter space\cite{footnote2}. Within each zone, 
$C_{\mu}$ is constant and the model is both stable and minimal. Outside this 
zone, the model is either too degenerate or overly complicated. For example, it 
will be degenerate (and $C_{\mu}$ will be too low) if $\tau$ is set too large. 
This is because equivalence classes will collapse into one another. Similarly, 
the model will be unnecessarily complicated (and $C_{\mu}$ will be too high) if 
$\tau$ is set too small. This is because distinctions will be made between words 
on the basis of insignificant differences in their profiles.

An analogy is the construction of a vocabulary for the structure of speciation 
of feline animals. If one is too fussy about the tail, Manx cats cannot be 
classed as domestic cats. If one's sole criterion is purring or a meow, a lion 
cub may be misclassed as a domestic cat. The correct classification of feline 
animals needs a finite amount of information to fall within the boundaries of a 
finite number of provisos.

In the case of Computational Mechanics we interpret effectively sofic models to 
be optimal. Thus we seek plateaus in the multi-dimensional parameter space. 
Generally, this space can contain many plateaus, the heights of which are the 
corresponding models' statistical complexities. If we want to forecast the data 
most accurately, we are looking for the highest plateau, which has the most 
stringent conditions\cite{footnote3}. More physically understandable models may 
exist on some lower plateaus where only the more dominant causal structures are 
preserved.

Thus, in the end, the success of the analysis depends upon the existence of 
effectively sofic plateaus of statistical complexity in the multi-dimensional 
parameter space and our ability to discover them. This is contingent upon the 
data that are supplied and how much computing power is available. It is 
important to bear in mind that the data are not only a function of the physical 
system's behaviour, but also of the measurement apparatus and any 
pre-processing. There are four main pitfalls (represented by corresponding model 
parameters):
\begin{enumerate}
\item Mischaracterization of the system by the measurement apparatus ($\ldots 
\chi, x$)
\item Degradation of data prior to the analysis by processing ($s$)
\item Insufficient data to resolve all structure present ($L$)
\item Insufficient computing power to resolve random structure ($n, \tau$)
\end{enumerate}
The apparatus may easily mischaracterize the system, either by introducing 
structure to the data which is foreign to the system's behaviour, or by 
neglecting to transcribe structure that should be present. This situation is 
most apparent
when the apparatus is clearly only taking measurements from a cross-section of 
the system. Nevertheless, if it is reasonable to assume in a particular case 
that the apparatus is capable of providing a good representation, then 
identified structure can be attributed to the system. In such cases we would 
also expect the statistical complexity to scale with the system's true 
complexity. Naturally, this is not valid when the cross-section happens to be an 
exact sub-system.

Although all processing degrades data, it may still be possible to correctly 
characterize all the structure present. This is because the degradation will 
usually produce noise, which can be ignored. A graver problem is when 
(uncharacterizable) noise represents some of the system's structure. The only 
solution may be to collect more data, but other preliminary approaches are to 
use a finer scale when coarse-graining and/or to digitise more finely. However, 
it is always necessary to choose sensible margins for the parameter search 
because some regions of the parameter space are computationally very costly to 
explore. For example, a trinary sequence is about seven hundred times as hard to 
fully analyse at a word length of eight than a binary sequence. You must have a 
good reason not to use binary.

An alternative approach may be useful when the data have a number of widely 
separated scales with structure; it may be more computationally efficient to 
construct higher-level equivalence classes than to persist with using longer and 
longer words. If this is the case there will be a change in the constant 
increase of $C_{\mu}$ with wordlength at the wordlength where the lower scale of 
structure is found, but the gradient will thereafter remain constant (until the 
next scale of structure is reached).

Classes on the next highest level are found by applying the same analysis method 
to the sequence expressed in terms of a set of primary level causal states for 
which $C_{\mu}$ has not yet converged. All information between the scales 
$n_{1}s$ and $n_{2}n_{1}s$ is lost in this process. Even so, it is a more 
preferable approach than simply further coarse-graining the data to intervals of 
$n_{1}s$ if one has reason to believe that the system's degrees of freedom at 
the two scales are coupled. The total statistical complexity is the sum of those 
calculated at each level, so it is in fact possible to test for such coupling by 
comparing the coarse-grained $C_{\mu}$ with the hierarchical value.

There is actually no reason why the prowords and epiwords should not come from 
different sequences, enabling the direct causal correlation of two systems, such 
as the solar wind and the magnetosphere.

\section{Conclusion}
\label{Conclusion}
Computational Mechanics is an intuitive and powerful way to study complicated 
linear outputs from physical systems. This is because the analysis identifies 
causal structure from data presented to it and constructs the minimal adequate 
model that fits these data. The information about this structure, and in 
particular its scales, can then be used to optimise more  physically plausible 
models. In this paper we have discussed in detail how the original formalism has 
to be used when applied to non-infinite sequences. The main conclusion is that 
models constructed by Computational Mechanics are good if, and only if, they are 
stable to the variation of the parameters used to construct them from the data. 
In addition, various more general postulates and definitions are made. These 
concern the general constructibility of models from a set of observations:
\begin{enumerate}
\item Structure which cannot be resolved from a set of data under any amount of 
computation is most usefully called {\em noise}
\item Structure which has not been resolved at a certain level of computation or 
memory, but which is resolvable from the set of data is usefully called {\em 
random}
\item Given enough data relevant to a system, there exists at least one 
connected region of finite size in the space of parameters used to construct 
models from those data within which an optimal model of the system is 
constructed.
\end{enumerate}

The prior undecidability of whether unresolved structure is noise or randomness 
is a direct parallelism of G\"{o}del's famous theorem. For a proof relating the 
two fields, but in a slightly different context, see 
G.J.~Chaitin\cite{Chaitin1974}.
\newline

The method developed in this paper was applied to magnetometer measurements of 
ionospheric currents for the years 1995, 1998 and 2000. The technique 
successfully constructed models, the simplest of which comprised a diurnal 
component and a Poisson-switched process with a timescale of about five hours 
that likely relates to the occurrence of magnetic substorms. The most 
complicated model could be used to forecast space weather.

A similar method was also proposed to characterize the causal relationship of 
any two systems, such as the solar wind and the magnetosphere.

 \appendix*
\section{Details of the geomagnetic data model}
Details of the simple stable model at wordlength
seven, $\tau \approx 0.14$. $x = 0.015$, at which
value 114 of 128 words are cut.\\

\begin{tabular}{|r|r|r|r|r|r|r|r|}
\hline
\multicolumn{1}{|c|}{Class} & \multicolumn{7}{|c|}{Equivalent surviving words} \\ \hline\hline
0 & 0 & 64 & 96 & 112 & 120 & 124 & 126 \\ \hline
1 & 1 & 3 & 7 & 15 & 31 & 63 \\ \cline{1-7}
2 & 127 \\ \cline{1-2}
\end{tabular}

\begin{tabular}{|r|r|r|}
\hline
\multicolumn{3}{|c|}{Transitions from Class 0} \\ \hline
Word Number (Class) & Word & Probability \\ \hline \hline
0 (0) & 0000000 & 0.481382 \\
1 (1) & 0000001 & 0.0759726 \\
3 (1) & 0000011 & 0.0729505 \\
7 (1) & 0000111 & 0.0641128 \\
15 (1) & 0001111 & 0.0540634 \\
31 (1) & 0011111 & 0.0546018 \\
63 (1) & 0111111 & 0.0532191 \\
64 (0) & 1000000 & 0.0338069 \\
96 (0) & 1100000 & 0.0234980 \\
127 (2) & 1111111 & 0.0514929 \\ \hline
\end{tabular}

\begin{tabular}{|r|r|r|}
\hline
\multicolumn{3}{|c|}{Transitions from Class 1}\\ \hline
Word Number (Class) & Word & Probability\\ \hline\hline
0 (0) & 0000000 & 0.0423272\\
64 (0) & 1000000 & 0.0291209\\
96 (0) & 1100000 & 0.0209377\\
112 (0) & 1110000 & 0.0209523\\
120 (0) & 1111000 & 0.0293505\\
124 (0) & 1111100 & 0.0442261\\
126 (0) & 1111110 & 0.0822315\\
127 (2) & 1111111 & 0.684994\\ \hline
\end{tabular}

\begin{tabular}{|r|r|r|}
\hline
\multicolumn{3}{|c|}{Transitions from Class 2}\\ \hline
Word Number (Class) & Word & Probability\\ \hline\hline
0 (0) & 0000000 & 0.0789801\\
64 (0) & 1000000 & 0.0907960\\
96 (0) & 1100000 & 0.0945274\\
112 (0) & 1110000 & 0.103234\\
120 (0) & 1111000 & 0.108831\\
124 (0) & 1111100 & 0.105721\\
126 (0) & 1111110 & 0.101990\\
127 (2) & 1111111 & 0.268035\\ \hline
\end{tabular}

\begin{tabular}{|r|r|r|}
\hline
\multicolumn{3}{|c|}{Average Transitions from Class 0}\\ \hline
Class & Average Word & Probability\\ \hline\hline
0 & [ 0.161, 0.102, 0.061, 0.040, 0.027, 0.009, 0.000 ] & 0.573587\\
1 & [ 0.000, 0.142, 0.288, 0.432, 0.603, 0.797, 1.000 ] & 0.374920\\
2 & [ 1.000, 1.000, 1.000, 1.000, 1.000, 1.000, 1.000 ] & 0.051493\\ \hline
\end{tabular}

\begin{tabular}{|r|r|r|}
\hline
\multicolumn{3}{|c|}{Average Transitions from Class 1}\\ \hline
Class & Average Word & Probability\\ \hline\hline
0 & [ 0.843, 0.735, 0.657, 0.579, 0.470, 0.306, 0.000 ] & 0.269146\\
1 & [ 0.000, 0.309, 0.444, 0.590, 0.714, 0.906, 1.000 ] & 0.045860\\
2 & [ 1.000, 1.000, 1.000, 1.000, 1.000, 1.000, 1.000 ] & 0.684994\\ \hline
\end{tabular}

\begin{tabular}{|r|r|r|}
\hline
\multicolumn{3}{|c|}{Average Transitions from Class 2}\\ \hline
Class & Average Word & Probability\\ \hline\hline
0 & [ 0.885, 0.752, 0.614, 0.463, 0.304, 0.149, 0.000 ] & 0.684080\\
1 & [ 0.000, 0.117, 0.221, 0.325, 0.532, 0.701, 1.000 ] & 0.047886\\
2 & [ 1.000, 1.000, 1.000, 1.000, 1.000, 1.000, 1.000 ] & 0.268035\\ \hline
\end{tabular}

\begin{tabular}{|r|r|}
\hline
Class & Prevalence\\ \hline\hline
0 & 0.483737\\
1 & 0.269561\\
2 & 0.246701\\ \hline
\end{tabular}

Statistical Complexity $=$ 1.51477 bits\\

\newpage

\begin{figure}[tbhp]
\caption{Two labelled diagraphs representing minimal models with statistical 
complexities of zero.}
\label{TwZrCmplxtDgrphs}
\end{figure}
\begin{figure}[tbhp]
\caption{A more complicated minimal model than that shown in 
Figure~\ref{TwZrCmplxtDgrphs}.}
\label{MrCmplctdDgrph}
\end{figure}
\begin{figure}[tbhp]
\caption{Statistical complexity of a noisy binary sequence, 5000 symbols long 
over a range of model construction parameters.}
\label{RndmSttstclCmplxt}
\end{figure}
\begin{figure}[tbhp]
\caption{Statistical complexity of a binary period 4 sequence, 5000 symbols 
long, 10\% flipped at random, over a range of model construction parameters.}
\label{10PrcntNsBnrPrd4SttstcclCmplxt}
\end{figure}
\begin{figure}[tbhp]
\caption{Analysis of the same sequence as in Fig.~3, with an assumed expletive 
prevalence of $x=0.075$.}
\label{10PrcntNsBnrPrd4SttstcclCmplxtX075}
\end{figure}
\begin{figure}[tbhp]
\caption{The minimal model of biased Poisson switches.}
\label{PssnSwtchDgrph}
\end{figure}
\begin{figure}[tbhp]
\caption{Two views of the variation of statistical complexity of the biased 
Poisson switch versus up-switching bias, $\alpha$ and down-switching bias, 
$\beta$.}
\label{PssnSwtchSttstcclCmplxtVrssBs1}
\end{figure}
\begin{figure}[tbhp]
\caption{Statistical complexity of a one million binary symbol sequence produced 
by a Poisson switch with $\alpha = \beta = 0.49$, over a range of model 
construction parameters.}
\label{AB49PssnSwtchSttstcclCmplxt}
\end{figure}
\begin{figure}[tbhp]
\caption{The minimal model of an unbiased switch which sustains for 4 symbols.}
\label{4SmblSstndSwtchDgrph}
\end{figure}
\begin{figure}[tbhp]
\caption{Statistical complexity of a one million binary symbol sequence produced 
by an unbiased switch which sustains for four symbols, over a range of model 
construction parameters.}
\label{4SmblSstndSwtchSttstcclCmplxt}
\end{figure}
\begin{figure}[tbhp]
\caption{Eastward D component of the magnetic deflection at Halley during 1998.}
\label{MgntcDflctn1998}
\end{figure}
\begin{figure}[tbhp]
\caption{Statistical complexity of a binary symbol sequence from about three 
years' worth of 180-minute time-averaged readings of the positive eastward 
component of the magnetic deflection at Halley, over a range of model 
construction parameters. The plateau at a word length of eight indicates major 
correlation at a period of 24 hours. All the plateaus in this diagram were 
stable to variation of the assumed expletive prevalence, here set at $0.4\%$.}
\label{180MntMgntcDflctnSttstcclCmplxt}
\end{figure}
\begin{figure}[tbhp]
\caption{Eastward D component of the magnetic deflection at Halley during 
February 1998.}
\label{MgntcDflctn199802}
\end{figure}
\begin{figure}[tbhp]
\caption{Eastward D component of the magnetic deflection at Halley during 16 
June 1998.}
\label{MgntcDflctn19980616}
\end{figure}
\begin{figure}[tbhp]
\caption{Statistical complexity of a 12011 binary symbol sequence from 40-minute 
time-averaged readings of the positive eastward component of the magnetic 
deflection at Halley, over a range of model construction parameters.}
\label{40MntMgntcDflctnSttstcclCmplxt}
\end{figure}
\begin{figure}[tbhp]
\caption{Predominant structure of the diagraph constructed from the Halley data 
at wordlength seven ($\tau = 0.14$, $x = 0.015$). Transitions with a probability 
less than 0.055 are not shown. The output labels are binarised averages, 
weighted according to the probabilities of individual words. This diagraph 
represents the least detailed structure discovered in the time series. More 
complicated diagraphs were constructed that are more suited to prediction than 
easy interpretation.}
\label{40MntMgntcDflctnDgrph}
\end{figure}
\begin{figure}[tbhp]
\caption{The main transitions of the minimal model for a binary period 36 
sequence at a wordlength of 7, $\tau = 0.14$, $x=0.015$.}
\label{Prd36Dgrph}
\end{figure}


\begin{thebibliography}{99}
\bibitem{footnote1} Even if the profiles are roughly equally occupied then a 
sequence length of about $L^{2n}$ is required to prevent statistical 
fluctuations from dominating the identification of equivalence classes.
\bibitem{footnote2} If we had been considering infinite sofic sequences this 
zone would be always be infinitely big because all structure would be resolved 
after a certain wordlength, and would continue to be resolvable. This does not 
usually apply to finite sofic sequences because beyond a certain threshold 
wordlength statistical fluctuations in the profiles are able to destroy the 
equivalence classes.
\bibitem{footnote3} Engineers would usually settle for that model constructed by 
the highest level of analysis, whether of not it is on a plateau. This is not 
acceptable from a scientific perspective because there is no way to justify that 
such models are not totally arbitrary.
\end{thebibliography}
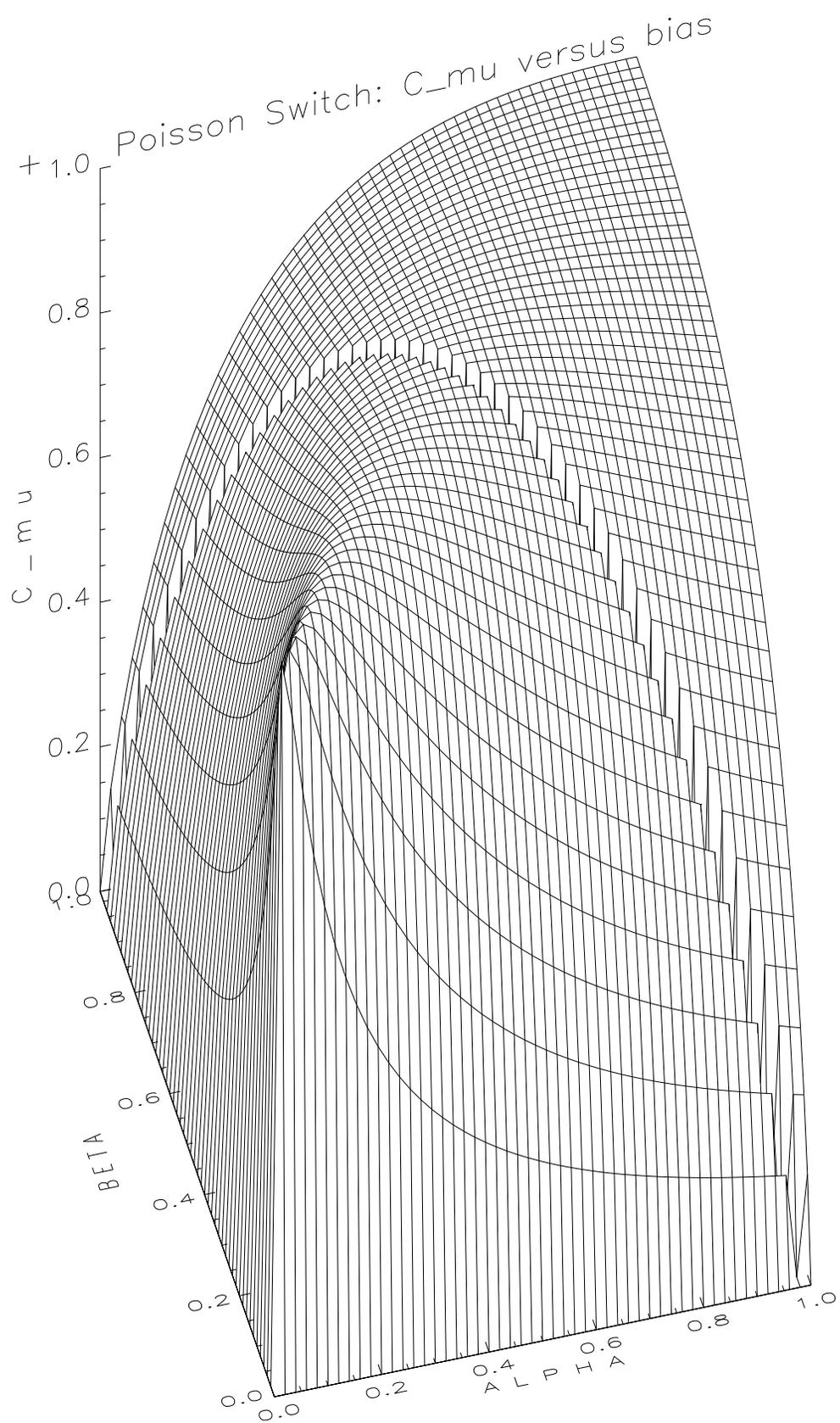
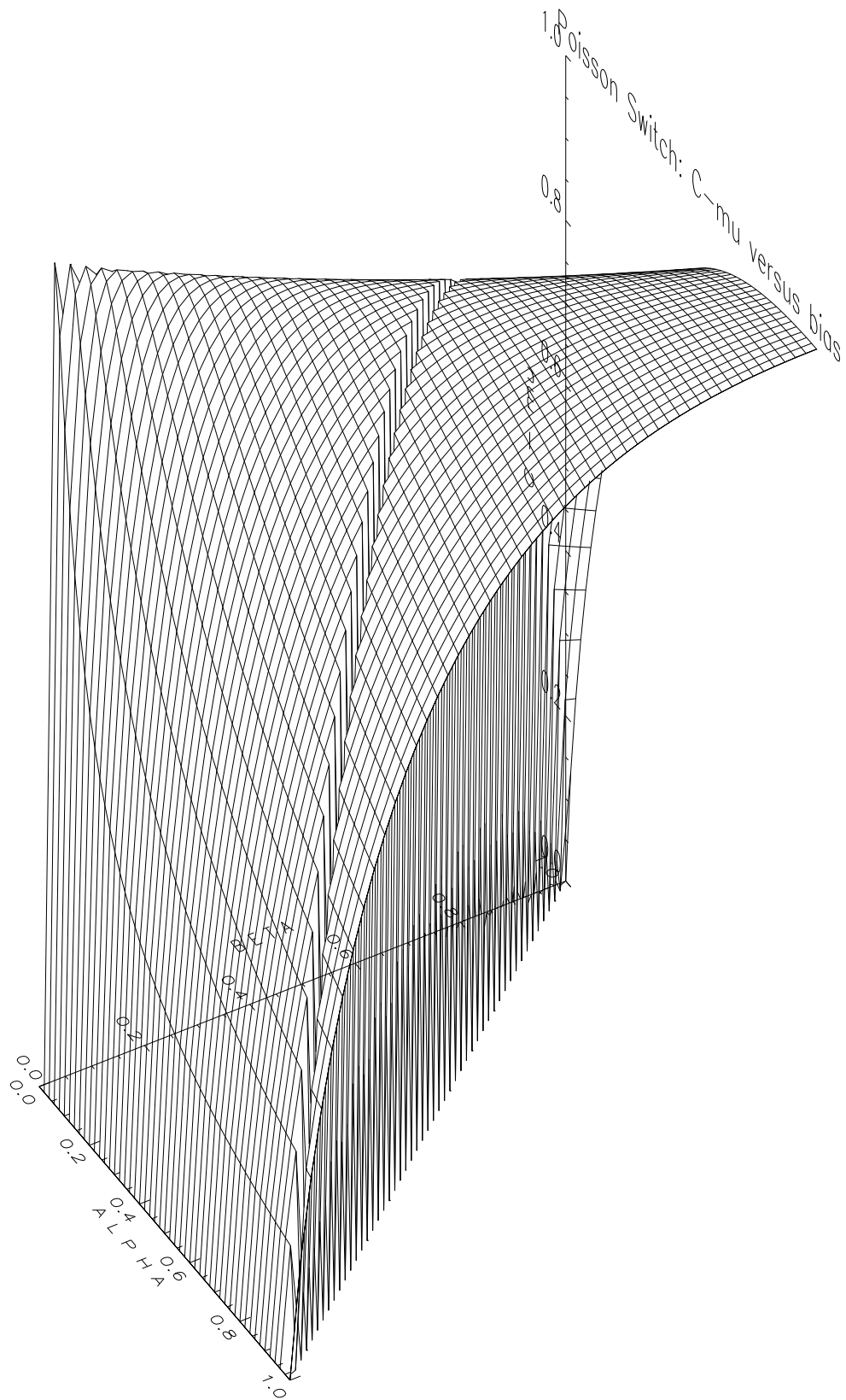
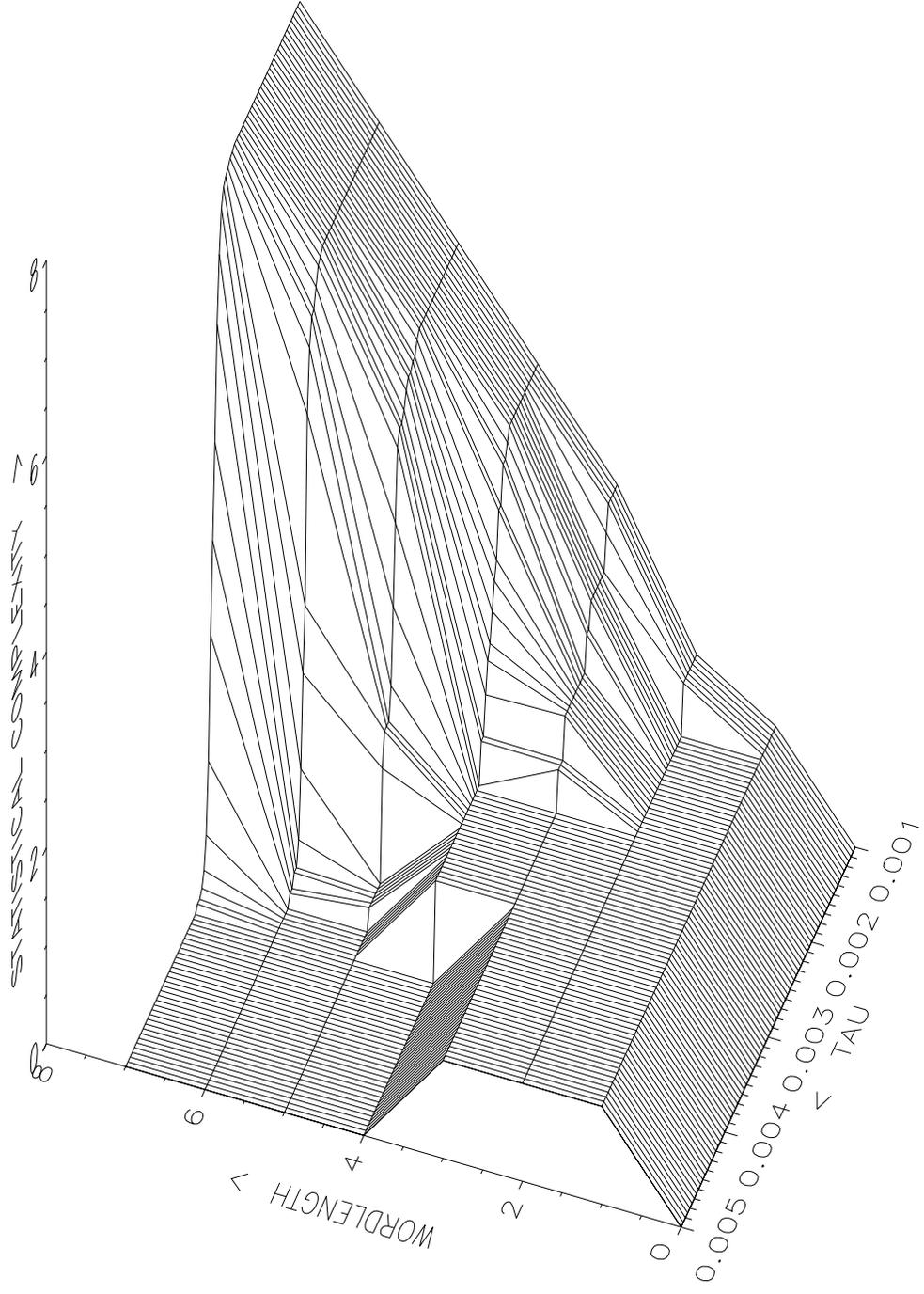
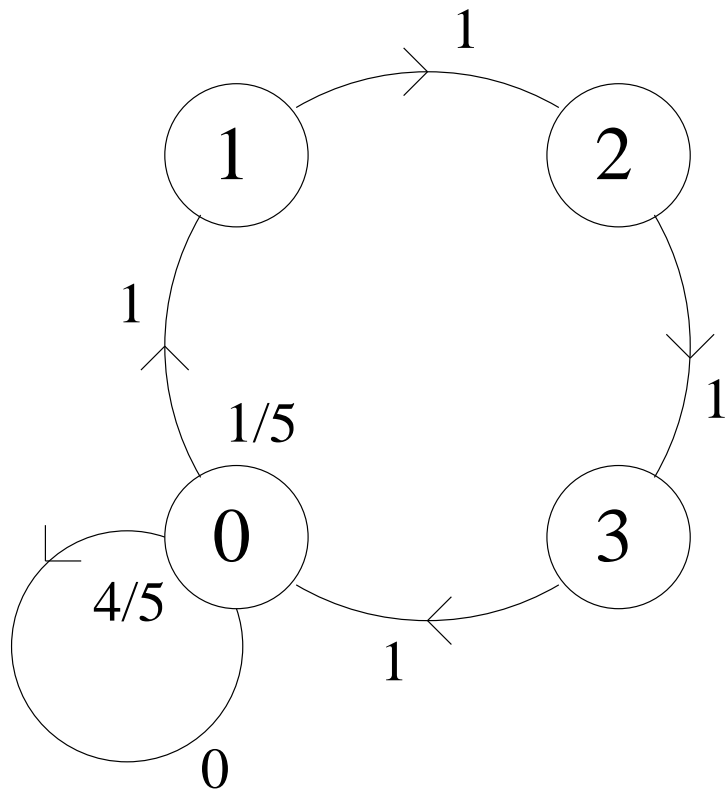
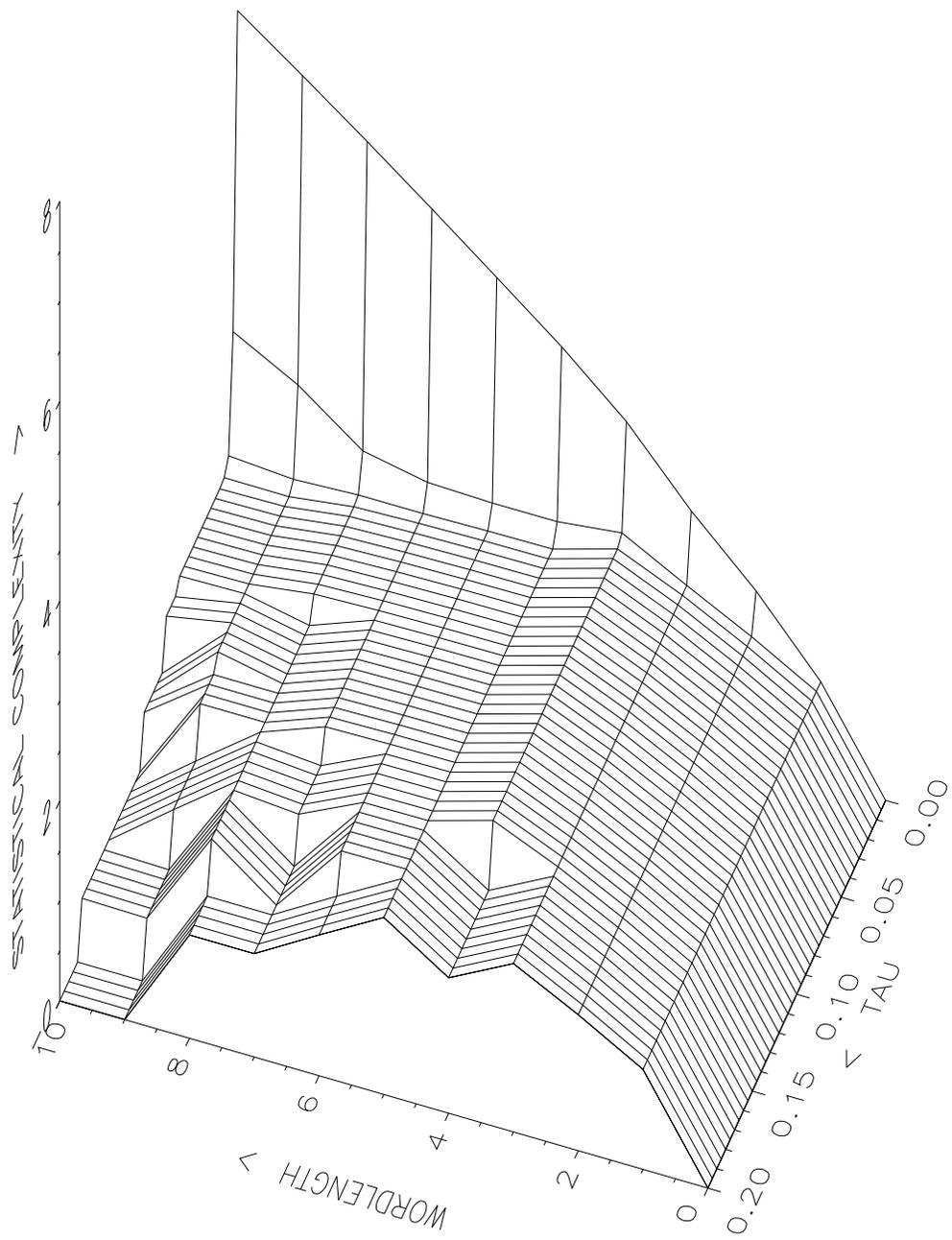
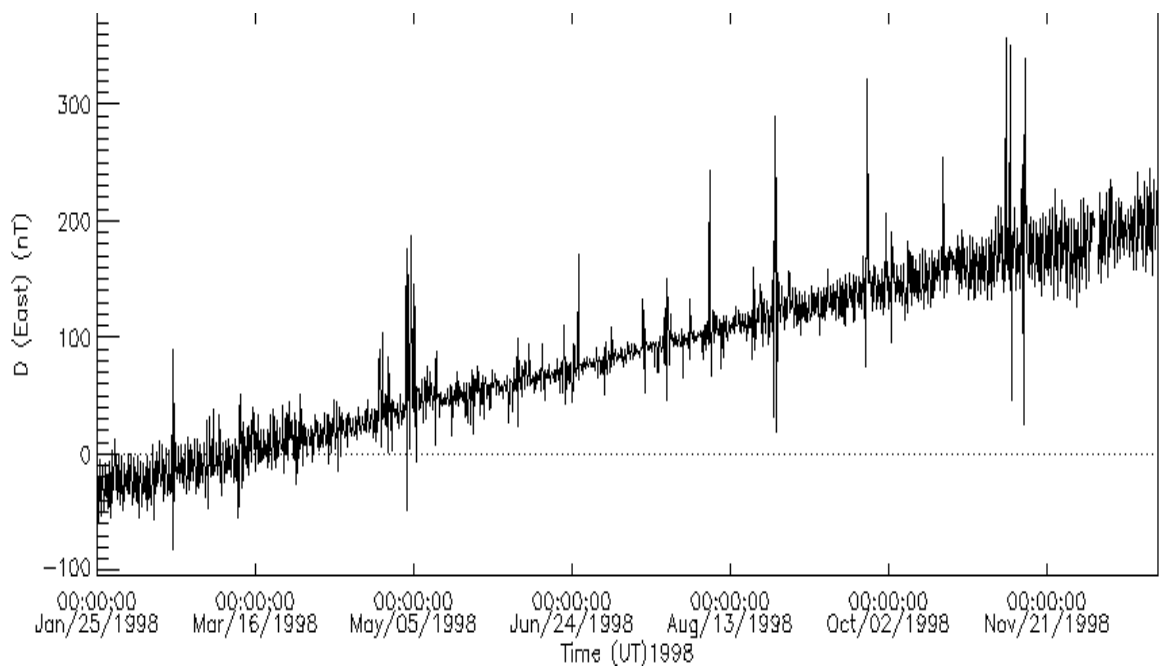
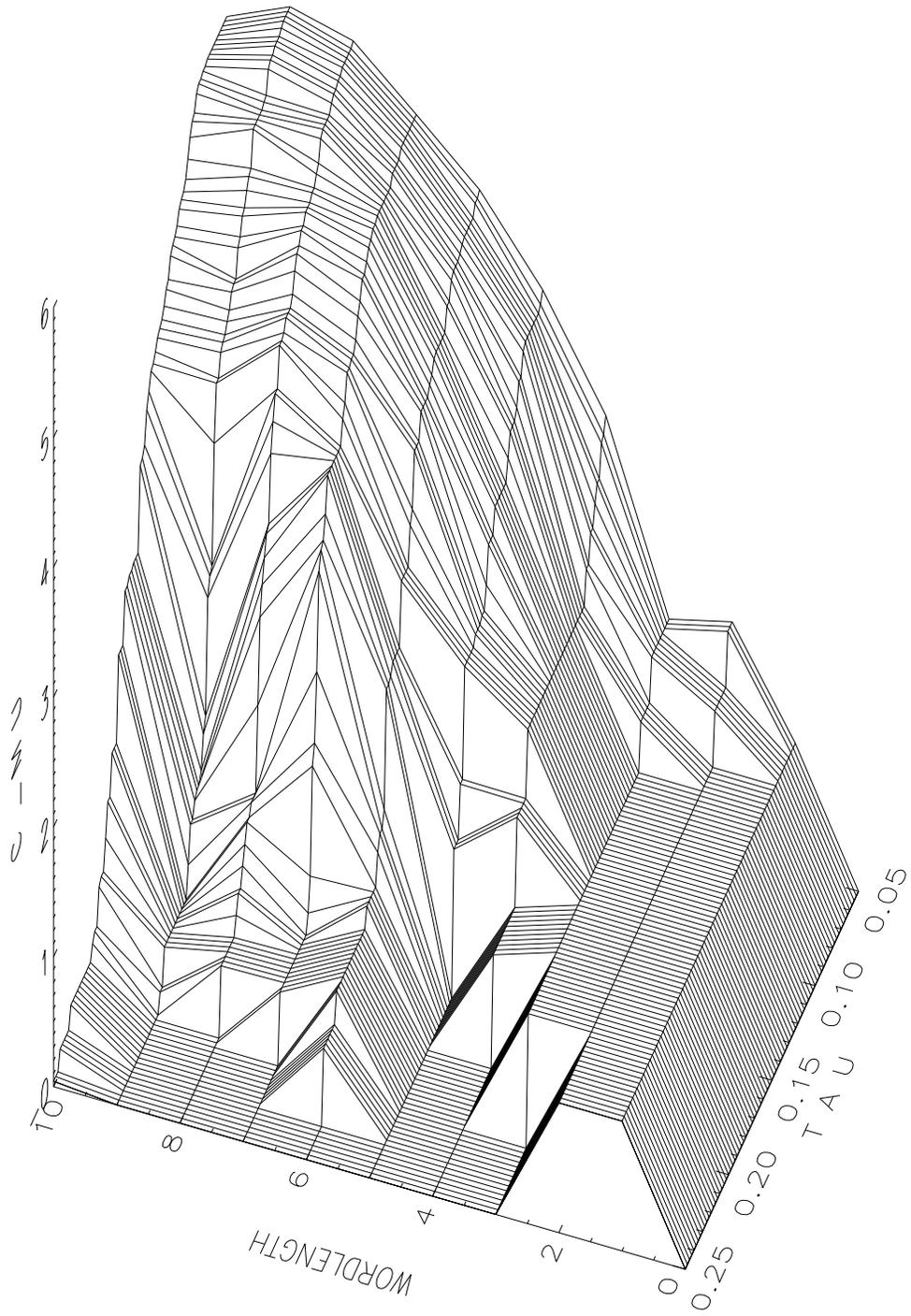
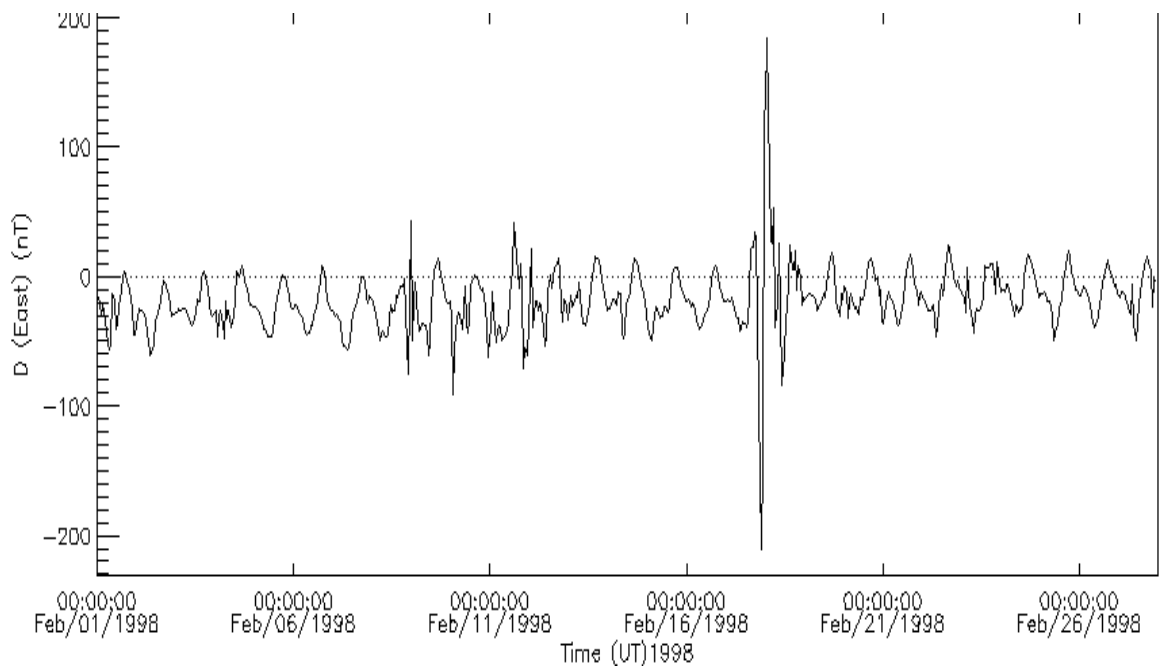
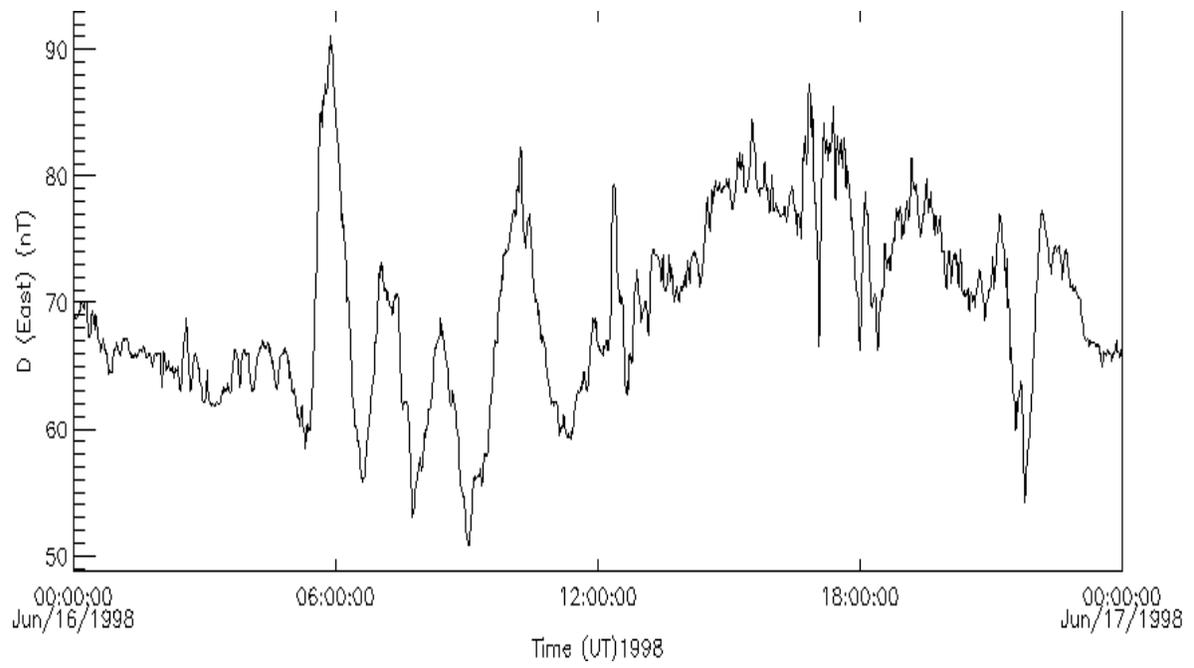
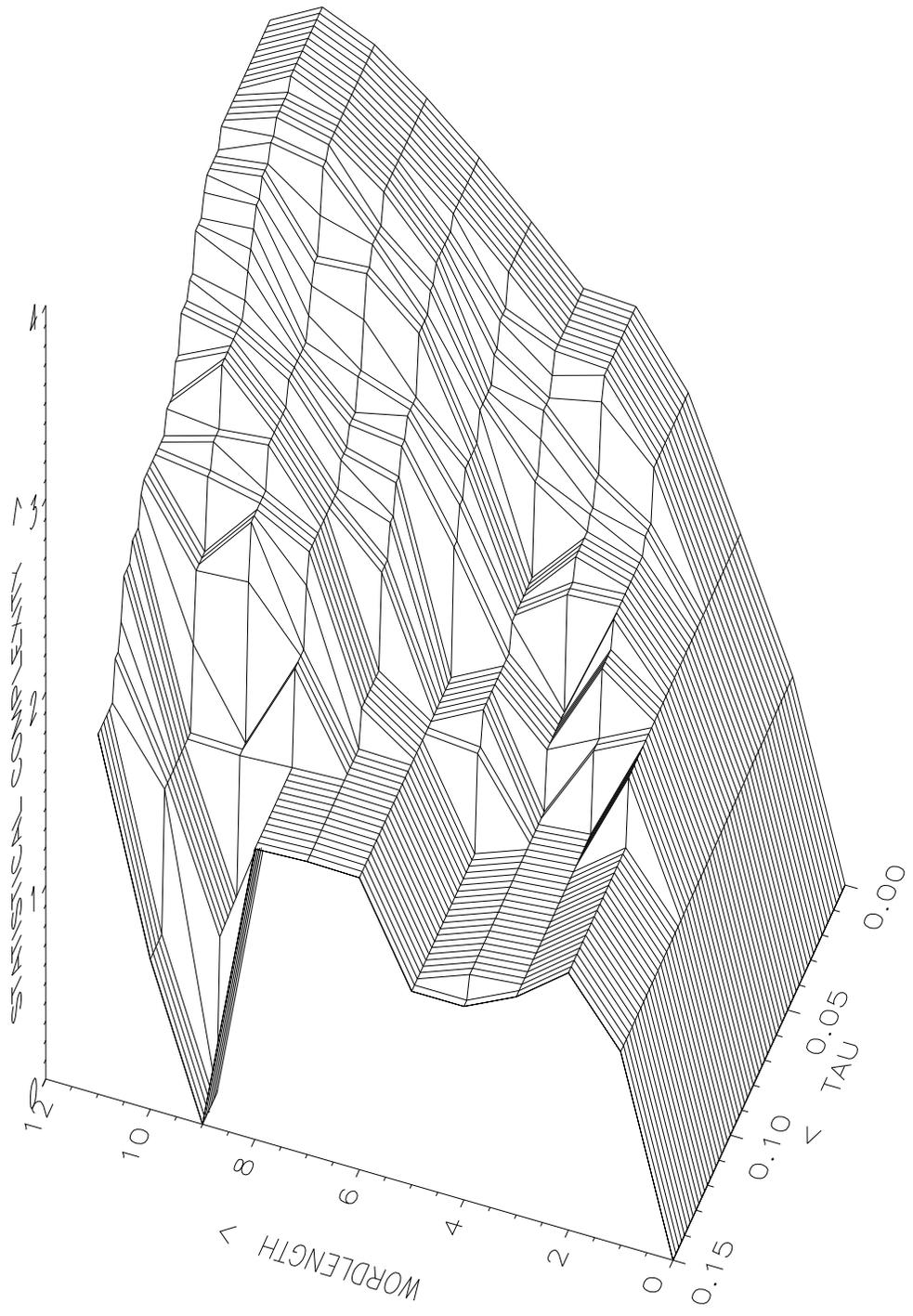
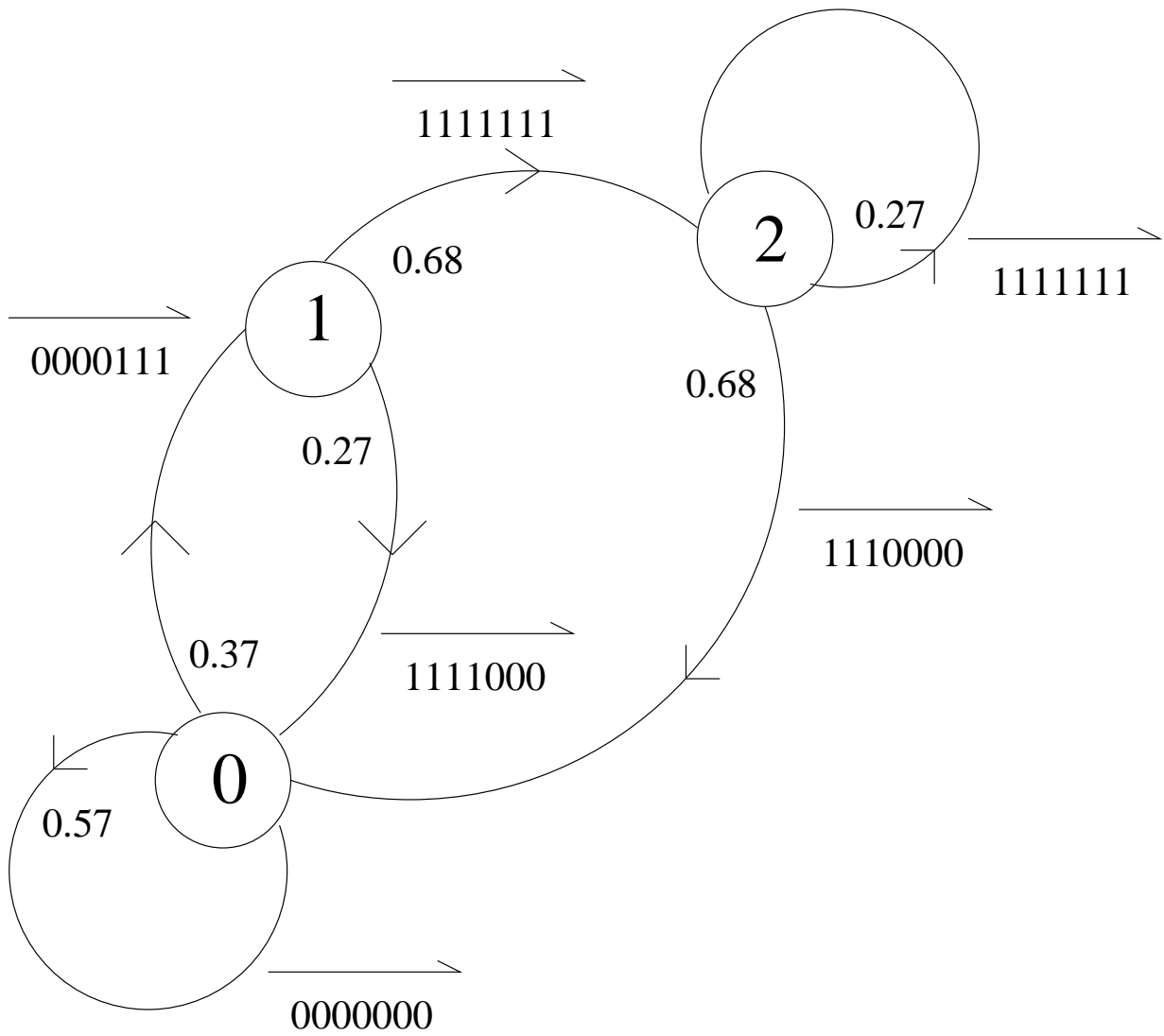
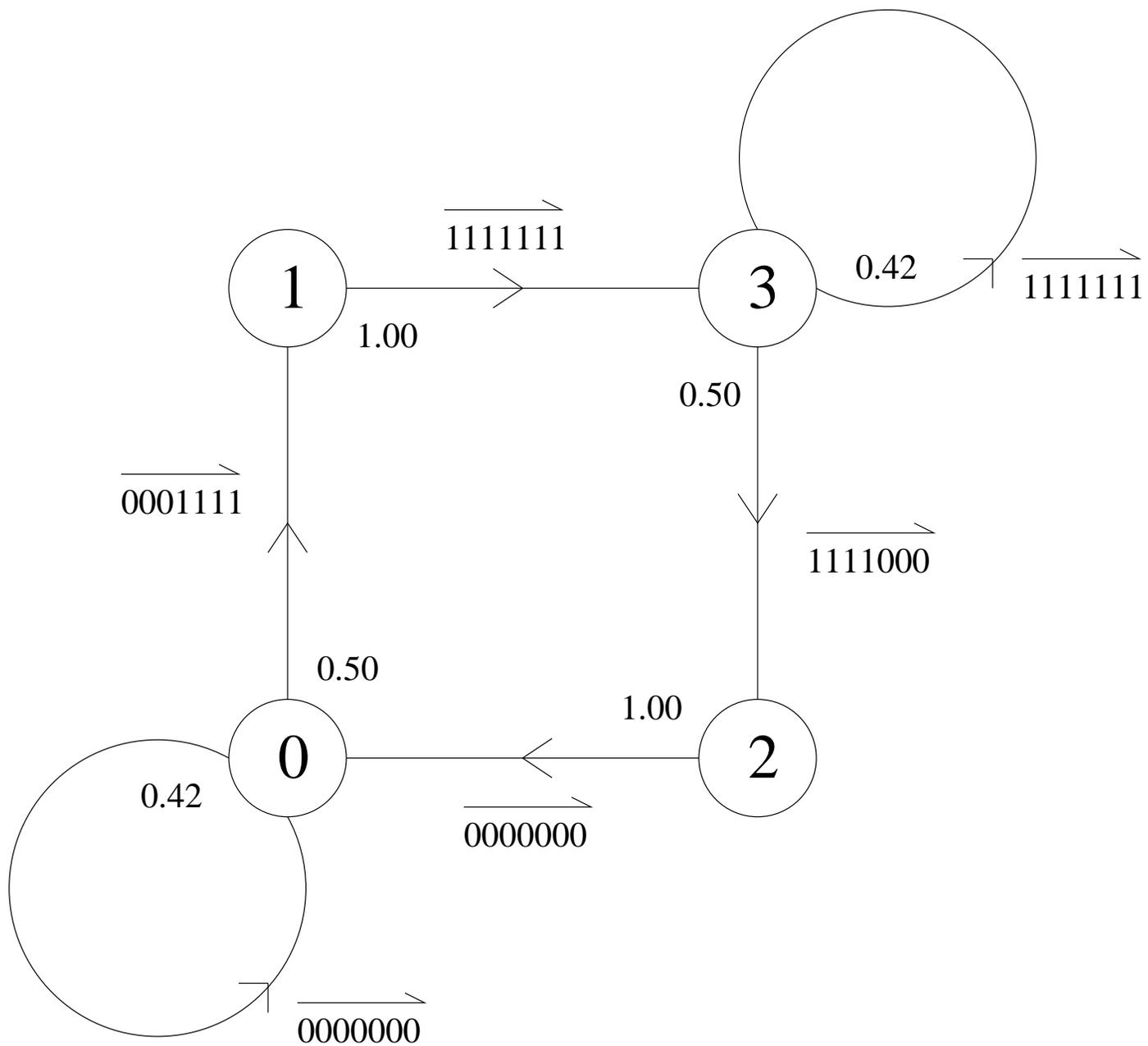
\end{document}